\documentclass[prb,twocolumn,showpacs]{revtex4}

\usepackage{amsmath,amsfonts,amssymb,bm}
\usepackage{dcolumn}
\usepackage[final]{graphicx}
\usepackage{bm}
\usepackage{times}

\begin{document}
\bibliographystyle{apsrev}

\title{A turnstile electron-spin entangler in semiconductors}

\author{Claudia Sifel}
\author{Ulrich Hohenester}\email{ulrich.hohenester@uni-graz.at}
\affiliation{Institut f\"ur Theoretische Physik,
  Karl--Franzens--Universit\"at Graz, Universit\"atsplatz 5,
  8010 Graz, Austria}

\date{May 21, 2003}

\begin{abstract}

We propose a single-electron doped quantum dot in a field-effect
structure as an optically triggered turnstile for spin-entangled
electrons. A short laser pulse excites a charged exciton, whose
quantum properties are transferred through tunneling and
relaxation to the spin entanglement between electrons in the dot
and contact. We identify the pertinent disentanglement mechanisms,
and discuss experimental detection and possible application
schemes.

\end{abstract}


\maketitle


Devices based on single quantum systems can provide single quanta.
This opens the possibility for the implementation of schemes based
on the fundamental laws of quantum mechanics, e.g., quantum
cryptography \cite{gisin:02} or quantum
computation.~\cite{bennett:00,bouwmeester:00} Within the field of
semiconductors it was soon realized that quantum
dots,~\cite{bimberg:98} sometimes referred to as {\em artificial
atoms},\/ are ideal candidates for such challenging future
applications, in particular in view of their high compatibility
with existing semiconductor technology. Indeed, in the seminal
work of G\'erard and Gayral~\cite{gerard:99} the authors proposed
a single quantum dot embedded in a microcavity as a viable
single-photon source; the applicability of this scheme was
demonstrated experimentally soon
after.~\cite{michler:00,santori:01,moreau:01} An important
technological improvement is due to Yuan {\em et
al.}~\cite{yuan:02} who succeeded to replace the optical
triggering by an electrical one.

A reversed approach was recently pursued by Zrenner {\em et
al.},~\cite{zrenner:02} where the authors used a quantum-dot
photodiode as an optically triggered {\em single-electron}\/
turnstile: a short laser-pulse coherently excites an exciton in a
quantum dot embedded in a field-effect structure; if the structure
is properly designed, such that tunneling occurs on a much shorter
timescale than radiative decay, the electron-hole excitation of
the quantum dot decays into a separated electron and hole within
the contacts, which is detected as the photocurrent. Within this
scheme it thus becomes possible {\em to transfer optical
excitations in a deterministic way to electrical currents}.

\begin{figure}[b]
\includegraphics[width=0.85\columnwidth,bb=115 320 445 645]{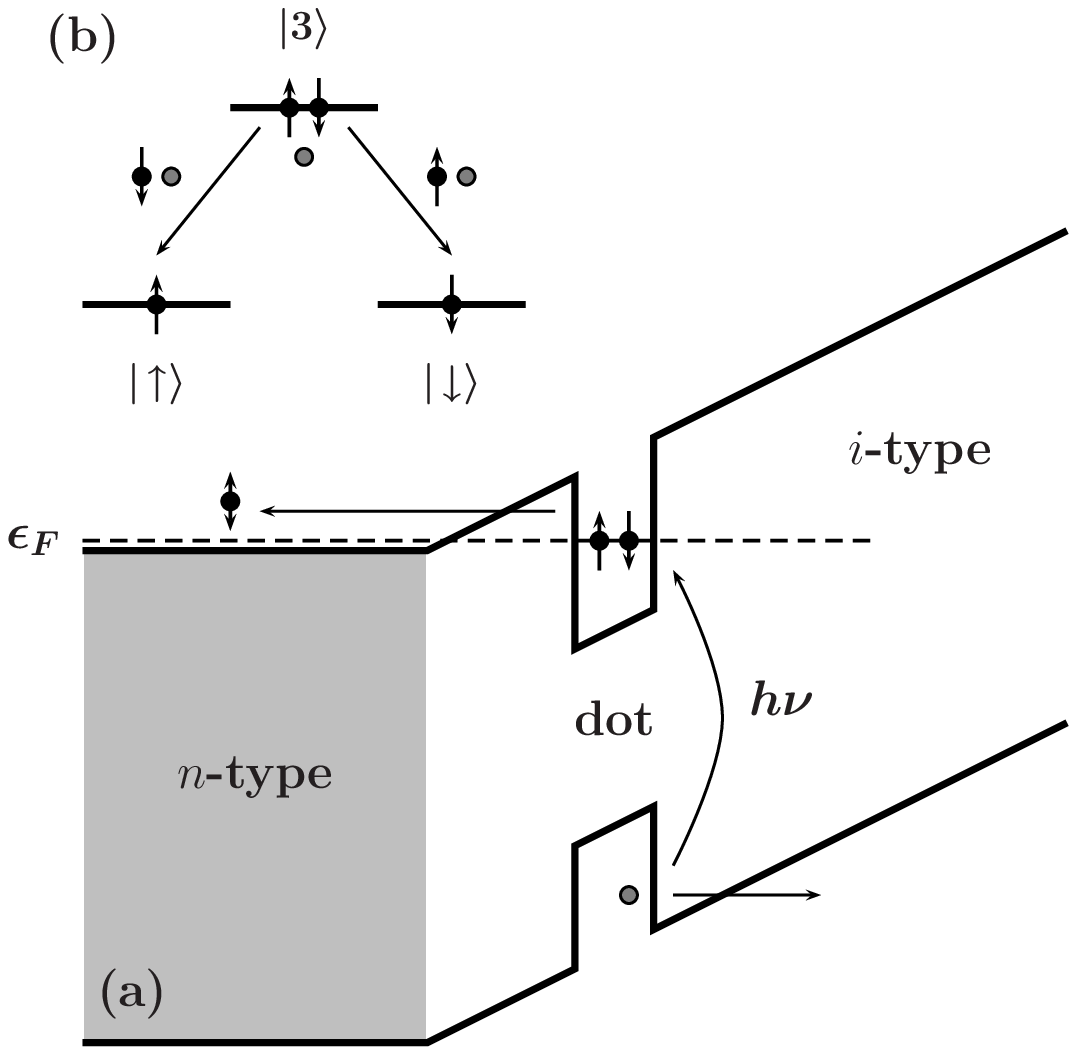}
\caption{(a) Schematic band diagram of the proposed structure. (b)
Level scheme of the spin-degenerate electron states
$|\sigma=\uparrow,\downarrow\rangle$ and the charged-exciton state
$|3\rangle$ in the dot.}
\end{figure}

In this paper we exploit this finding to propose a device which
allows the optically triggered creation of a spin-entangled
electron pair. The proposed structure (Fig.~1a) is identical to
the one used by Zrenner {\em et al.},~\cite{zrenner:02,beham:01}
with the only exception that the dot is initially populated by a
single surplus electron; this can be achieved by applying an
external bias voltage such that an electron is transferred from a
nearby $n$-type reservoir to the
dot,~\cite{warburton:00,findeis.prb:01} where further charging is
prohibited because of the Coulomb blockade. Optical excitation of
this structure then results in the excitation of a {\em charged
exciton},\/ i.e., a complex consisting of two electrons and a
single hole;~\cite{warburton:00,hartmann.prl:00,findeis.prb:01}
appropriate tuning of light polarization and frequency allows to
selectively excite the charged-exciton groundstate, where the two
electrons have opposite spin orientations. Since within the
field-effect structure the charged exciton is not a stable
configuration, in a consequent step one electron and hole will
tunnel out from the dot to the nearby contacts; here, the system
can follow two pathways, where either the electron in the dot has
spin-up and the one in the reservoir spin-down orientation or vice
versa. According to the laws of quantum mechanics, the total state
of the system thus becomes a superposition of these two
configurations; as will be proven below, {\em in this state the
electron spins are maximally entangled}.\/ Thus, the proposed
device is an optically triggered turnstile for spin-entangled
electrons, which could be used in future quantum information
applications to establish entanglement between spatially separated
sites.

In a sense, our scheme is similar to the proposal of Benson {\em
et al.}~\cite{benson:00} in which entangled photons are created in
the cascade decay of a biexciton. However, in the system of our
present concern additional difficulties arise because the
tunnel-generated electron and hole do not propagate freely (as
photons would in the corresponding scheme) but are subject to
interactions in the contact. The resulting scatterings of the
entangled particles hamper a straightforward interpretation of the
functionality of the proposed device and call for a careful
theoretical analysis. It is the purpose of this paper to provide a
comprehensible theory accounting for the complete cascade process
of: the buildup of three-particle coherence through tunneling; the
swapping of quantum coherence to spin entanglement through
dephasing and relaxation in the reservoirs; and finally the
process of disentanglement through spin-selective scatterings.
Since the main emphasis of our work is on the identification of
the basic schemes underlying the buildup and decay of
entanglement, we rely on a simplified description scheme of
environment interactions, which will allow us to derive analytic
expressions throughout.


Our model system comprises (Fig.~1b): the spin-degenerate electron
groundstates $|\sigma\rangle$ and the charged-exciton groundstate
$|3\rangle$ in the dot (with energies $E_\sigma$ and $E_3$,
respectively); the electron and hole states in the reservoir,
described by the usual field operators $c_{k\sigma}$ and $d_{k'}$
(energies $\epsilon_{k\sigma}^e$ and $\epsilon_{k'}^h$) with $k$
labeling the quantum numbers (e.g., wavevector and band index),
and we have explicitly indicated the electron spin. The
Hamiltonian of the system without interactions $H_o$ thus reads

\begin{equation}
  H_o=\sum_{\sigma}E_\sigma |\sigma\rangle\langle\sigma|+
  E_3|3\rangle\langle 3|+
  \sum_{k\sigma}\epsilon_{k\sigma}^e c_{k\sigma}^\dagger c_{k\sigma}+
  \sum_{k'}\epsilon_{k'}^h d_{k'}^\dagger d_{k'}.
\end{equation}

Since we are dealing with an {\em open system}\/ (i.e., system
interacting with its environment) we have to adopt a
density-matrix description.~\cite{walls:95,scully:97} Let us
assume that initially the electron spin direction is undetermined,
i.e., the corresponding density matrix is a mixture $\rho=\frac 1
2 \sum_\sigma |\sigma\rangle\langle\sigma|$. When at time $t=0$
the dot is subject to an unpolarized optical
$\pi$-pulse~\cite{zrenner:02} it will be excited to state $3$.
Hence, the initial density matrix is $|3\rangle\langle 3|$
(although the proposed scheme would also work for charged-exciton
occupancies less than one, as discussed below).

For the system's time evolution we employ a master-equation
framework of Lindblad form~\cite{walls:95,scully:97}

\begin{equation}\label{eq:lindblad}
  \dot\rho=-i[H_o,\rho]-
  \frac 1 2\sum_i(L_i^\dagger L_i\rho+\rho L_i^\dagger L_i)+
  \sum_i L_i\rho L_i^\dagger,
\end{equation}

\noindent within which scatterings are described in the usual
Markov and adiabatic approximations. In Eq.~\eqref{eq:lindblad}
the $L_i$'s are the Lindblad operators which account for the
different scattering channels.

{\em Tunneling.}---For low temperatures and early times we can
safely neglect phonon processes and radiative decay in the dot,
and tunneling becomes the only relevant scattering channel. Quite
generally, the question whether combined electron-hole tunneling
dominates over separate tunneling (as we will assume) depends on
the design of the structure. In Ref.~\onlinecite{beham:01} the 
authors measured tunneling lifetimes between 29 and 330 ps, where
the exact value strongly depends on the internal electric 
field;~\cite{larkin:03} alternatively, it might be advantegeous
to use type-II quantum dots~\cite{janssens:02} where the hole is
only Coulomb bound and much shorter tunneling lifetimes could
be achieved. However, such details are not crucial to
our study and the only relevant assumptions are: first, all
tunneling processes are independent of spin; second, since the
hole enters with a high excess energy into the contact it
immediately suffers an inelastic scattering, which guarantees that
tunneling is an irreversible process. In our calculations the
latter point is taken into account by tracing over the hole
degrees of freedom in the reservoir and neglecting terms
$tr_h\rho\;d^\dagger d$. Within this framework and assuming
tunneling matrix elements $\hat t$ independent of $k$ and $k'$, we
can solve Eq.~\eqref{eq:lindblad} through an {\em unraveling of
the master equation}~\cite{plenio:98} to obtain:~\cite{sifel:03}

\begin{equation}\label{eq:conditional}
  \rho(t)\cong e^{-\lambda t}|3\rangle\langle 3|+\lambda\int_0^t
  dt'\; e^{-\lambda t'} U(t,t')
  |\Psi_1\rangle\langle\Psi_1|U(t',t),
\end{equation}

\noindent with: $\lambda=2\pi|\hat t|^2\sum_\sigma \int d\omega_e
d\omega_h\; g_e(\omega_e) g_h(\omega_h) \delta(\omega_e +\omega_h
+E_\sigma-E_3)$ the total tunneling rate; $g_{e,h}(\omega)$ the
electron and hole density-of-states in the reservoir; $U(t,t')$
the time evolution operator in the reservoir; and
$|\Psi_1\rangle\langle\Psi_1|$ the density matrix after tunneling.
In the spirit of the quantum-jump approach,~\cite{plenio:98} in
Eq.~\eqref{eq:conditional} the first term can be interpreted as
the {\em conditional density matrix}\/ for no tunneling (which
decays with $e^{-\lambda t}$) whereas the second term is the
conditional evolution after tunneling. The corresponding density
matrix is obtained from~\cite{plenio:98} $L_i|3\rangle\langle
3|L_i^\dagger/tr(.)$, where $L_i$ is the Lindblad operator for
tunneling and the denominator ensures $tr\;\rho=1$, which gives:

\begin{equation}\label{eq:psi}
  |\Psi_1\rangle\propto\sum_\sigma
  \int_{\epsilon_F}^{\omega_c}d\omega\;C_\sigma^\dagger(\omega)
  |\bar\sigma\rangle.
\end{equation}

\noindent Here, $\epsilon_F$ is the Fermi energy of the $n$-type
reservoir; $\omega_c$ is a cutoff energy due to the kinematics of
the tunneling process; $C_\sigma(\omega)=\sum_k c_{k\sigma}
\delta(\omega-\epsilon_{k\sigma}^e)$; and $\bar\sigma$ a
spin-orientation antiparallel to $\sigma$. Eq.~\eqref{eq:psi} is
an important and non-trivial result. First, it demonstrates that
despite the incoherent nature of tunneling and hole relaxation the
electron system can be described in terms of wavefunctions; we
note here in passing that the detection of the hole would even
allow to purify this wavefunction,~\cite{bouwmeester:00} which
might be of relevance when initially $\rho$ is not equal to
$|3\rangle\langle 3|$. Second, a closer inspection of
Eq.~\eqref{eq:psi} reveals that the spin part
$C_\uparrow^\dagger|\!\downarrow\rangle+
C_\downarrow^\dagger|\!\uparrow\rangle$ is a {\em maximally
entangled state}\/ of the electrons in the dot and reservoir. We
emphasize that this maximal entanglement is independent of the
spin basis, which guarantees that our scheme is not deteriorated
by possible polarization anisotropies of the dot states
(fine-structure splittings).

{\em Dephasing and relaxation.}---After tunneling the system
propagates in presence of scatterings, as described by $U(t,t')$
in Eq.~\eqref{eq:conditional}. Quite generally, we assume that the
orbital degrees of the reservoir electron are subject to much
stronger interaction channels (e.g., phonons) than the spin
degrees, as evidenced by the long measured spin lifetimes
($\sim$ns) in $n$-doped semiconductors.~\cite{kikkawa:98} For that
reason, let us first consider an elastic electron scattering which
does not depend on spin, i.e., Lindblad operators of the form
$\Gamma^\frac 1 2 \sum_\sigma C_\sigma^\dagger(\omega)
C_\sigma(\omega)$ with $\Gamma$ the scattering rate. Unraveling
the corresponding master equation in an analogous fashion to
Eq.~\eqref{eq:conditional}, we again recover a conditional
evolution for no scattering (which decays with $e^{-\Gamma t}$)
and a remainder which describes the effects of scattering; here,
the density matrix after scattering becomes:~\cite{sifel:03}

\begin{equation}\label{eq:rho}
  |\Psi_1\rangle\langle\Psi_1|\longrightarrow
  \sum_{\sigma\sigma'}
  \int_{\epsilon_F}^{\omega_c}d\omega\;
  C_\sigma^\dagger(\omega)
  |\bar\sigma\rangle\langle\bar\sigma'|C_{\sigma'}(\omega).
\end{equation}

\noindent In comparison to Eq.~\eqref{eq:psi} the density matrix
of Eq.~\eqref{eq:rho} is diagonal in $\omega$, i.e., the elastic
scattering has led to a destruction of the phase coherence (i.e.,
dephasing). However, {\em the spin part still shows the same
degree of entanglement},\/ where similar conclusions would apply
for inelastic but spin-independent scatterings. {\em Thus, the
decay of an optically excited charged-exciton indeed generates a
robust spin entanglement between the electron in the dot and
reservoir.}

{\em Disentanglement.}---We finally comment on the process of
disentanglement. In fact, any scattering channel which couples
with unequal strength to the spins (or affects only one spin
orientation) is responsible for such entanglement decay. Naively,
one could expect that a preferential scattering of, e.g., spin-up
electrons in the reservoir would establish a stronger degree of
spin-down population in the dot; however, this is not supported by
our calculations which show that any spin-selective 
scattering~\cite{comment:lindblad}
forces the spins {\em with equal probability}\/ to one of the two
orientations. Thus, to experimentally detect spin entanglement in
the proposed scheme both electrons have to be monitored. This
could be achieved by introducing a ferromagnetic contact at the
interface of the $n$-doped region, which acts as a spin filter for
the reservoir electron. Transmission across the interface
corresponds to a spin measurement which also determines the spin
orientation $\sigma$ of the electron in the quantum dot. The
resulting state $|\sigma\rangle$ could be probed by a second,
time-delayed optical $\pi$-pulse whose polarization is chosen such
that it selectively excites the $\sigma$--3 transition. Thus, the
transmission into the ferromagnet is accompanied by the optical
excitation of a second charged exciton in the dot (which
consecutively is transferred to an electric current). On the other
hand, if the second optical pulse arrives before disentanglement,
the dot density matrix is a mixture $\frac 1 2\sum_\sigma
|\sigma\rangle\langle\sigma|$ and optical excitation occurs only
with a 50\%-probability, which results in a distinctly different
noise characteristics of the photocurrent.


In conclusion, we have proposed a scheme for an optically
triggered spin entanglement of electrons in semiconductors. It
consists of a single-electron doped quantum dot embedded in a
field-effect structure. Optical excitation of an additional
electron-hole pair (charged exciton) is transferred through
tunneling to a photocurrent, where the spins of the electrons in
the dot and reservoir are maximally entangled. We have discussed
that this entanglement is robust against dephasing and relaxation
processes which are not spin-selective, and thus benefits from the
long spin lifetimes in semiconductors. The proposed device might
be useful in future quantum information applications to establish
entanglement between spatially separated sites; there, it might be
advantageous to replace the $n$-type reservoir by quantum wires
(for a natural realization of such combined dot-wires structures
see, e.g., Ref.~\onlinecite{hartmann.prl:00}). Finally, in
contrast to other proposal for spin entanglement in
semiconductors~\cite{engel:01} our scheme allows the creation of
spin-entangled electrons on demand (through optical triggering).

U. H. is grateful to Jaroslav Fabian for helpful discussions. This
work has been supported in part by the Austrian science fund FWF
under project P15752.


\begin{thebibliography}{24}
\bibitem[{com()}]{comment:lindblad}
\bibinfo{note}{In our calculations we use Lindblad operators of the form
  $L=\sqrt\Gamma(\cos^2\theta\,C_\uparrow^\dagger C_\uparrow+
  \sin^2\theta\,C_\downarrow^\dagger C_\downarrow)$, with $\Gamma$ the
  scattering rate and $\theta$ an angle determining the degree of asymmetry.}
\end{thebibliography}
\end{document}